\documentstyle[titlepage,12pt]{article}
\begin{document}
\parindent 2em

\begin{titlepage}
\begin{center}
\vspace{12mm}
{\LARGE Superconductor-insulator quantum critical point in 
$1+\epsilon$ dimensions}

\vspace{15mm}

Igor F. Herbut$^{*}$\\

Department of Physics and Astronomy, University of British Columbia, 
6224 Agricultural Road, Vancouver B. C., Canada V6T 1Z1\\

\end{center}
\vspace{10mm}
{\bf Abstract:} A system of spinless fermions in $d=1+\epsilon$ 
dimensions, at zero-temperature 
 and in random external potential is studied using the 
perturbative renormalization group to first order in disorder and 
to second order in interaction. We find a superconductor-to-Anderson 
insulator quantum fixed point at an infinitesimal value of disorder
and calculate the correlation length and the dynamical exponents  
to the lowest order in $\epsilon$ and 
in interaction. The scaling 
of conductivity with temperature and the behavior of characteristic 
temperature scales on both sides of the transition is determined. 
The model may have relevance for a p-wave superconductor at 
low temperatures in strongly disordered media. 


\vspace{80mm}

$^{*}$ E-mail: herbut@theory.physics.ubc.ca
\end{titlepage}


The effect of Anderson localization on superconductivity or superfluidity 
represents a fundamental but still incompletely understood problem in 
condensed matter physics. 
The effects of disorder may be expected to be 
particularly pronounced in low dimensions, $d\leq 2$, where 
arbitrarily weak disorder localizes all 
single-particle states \cite{abrahams} in a non-interacting system. 
Experiments on 
thin, microscopically disordered films  at 
low temperatures, for example,  have revealed 
a sharp separation between the 
insulating and superconducting regimes as 
some parameter controlling the amount of disorder is varied 
\cite{haviland}. These measurements 
are usually interpreted as  consequences of 
the proximity of a  
quantum ($T=0$), superconductor-insulator (SI) 
critical point \cite{weichman}, \cite{fisher}. Properties of 
such a critical point have  
been a subject of intense investigations in recent years \cite{wallin},
particularly in the context of the theory of superfluid-Bose glass 
transition, which assumes that the important degrees of freedom close to the 
transition are bosonic, preformed Cooper pairs.  
One of the  obstacles in better quantitative 
understanding of the quantum SI transition 
lies in the fact that the assumed SI fixed point 
resides in the region of strong disorder,  
and is not easily revealed in perturbative calculations 
\cite{weichman}, \cite{halperin}. 
A notable  exemption from this rule 
is the SI transition in one dimension \cite{apel}, 
\cite{giamarchi}. At $T=0$ and without disorder  
$d=1$ is the lower critical dimension for superconductivity; only  
power-law superconducting correlations are possible since the 
true long-range superconducting 
order is eliminated according to Coleman-Mermin-Wagner theorem. 
In the language of 
renormalization group, this is manifested in 
scaling towards finite attractive fixed 
points \cite{solyom}. 
When disorder is added, an immediate consequence of the special 
situation in $d=1$ is that the SI fixed  point is located at zero-disorder 
(and at finite interaction),  
suggesting an   approach to the problem of the 
SI transition in higher dimensions 
via expansion around $d=1$. Such an expansion would be of 
particular value since there may not be a finite upper critical 
dimension for the problem \cite{weichman}.

  In this Letter we address some of the above issues by studying a 
model of interacting, disordered spinless fermions in $d=1+\epsilon$. 
The model may be appropriate for the description of 
the SI transition from a superconductor 
state with p-wave symmetry. Without disorder and for $\epsilon>0$, the
 interaction coupling constant is 
irrelevant in infrared if positive, and runs towards $-\infty$ if negative, 
indicating the formation of 
superconducting long-range order. 
In the  presence of disorder, we find a 
fixed point that controls the quantum SI transition 
at the value of dimensionless resistance of order $\epsilon$, 
and of attractive interaction of order unity (see Figure 1). 
  The critical 
exponents that characterize the fixed point are evaluated to the lowest 
order in $\epsilon$ and perturbatively in strength of interaction. 
The correlation length exponent
 diverges as $\epsilon\rightarrow 0$, and the dynamical exponent 
acquires a positive correction of order $\epsilon$. 
The fact that the fixed point is infinitesimal in disorder 
enables one to obtain the low-temperature scaling of conductivity 
with temperature in  simple way. 
Due to the pair-breaking nature of the disorder in the problem,
superconducting gap vanishes at the critical point. 

We consider a system of spinless fermions in
 $d$-dimensions, $1\leq d \leq 2$,  in random external 
potential distributed according to the probability: 
\begin{equation}
P[V(\vec{r})]\propto \exp -\int d^{d}\vec{r} d^{d}\vec{r'} V(\vec{r})
W^{-1}(\vec{r}-\vec{r'}) V(\vec{r'}). 
\end{equation}
The quantity of interest is the 
zero-temperature free energy $G=-\langle \ln Z\rangle$, where brackets 
denote averaging over disorder configurations. Utilizing the standard 
replica trick \cite{edwards}, after averaging over disorder we arrive at 
the zero-temperature action for fermions:
\begin{equation}
S=S_{0}+S_{int}+S_{dis}
\end{equation} 
where, 
\begin{eqnarray*}
S_0 =\sum_{\alpha=1}^{N}
\int d^{d}\vec{K} 
\frac{d\omega}{2\pi}
(-i\omega+e(\vec{K}))\Psi_{\alpha}^{\dagger}(\vec{K},\omega) 
\Psi_{\alpha}(\vec{K},\omega), 
\end{eqnarray*}
\begin{eqnarray*}
S_{int}=\sum_{\alpha=1}^{N} \int 
\prod_{j=1}^{4} \frac{d^{d}\vec{K}_{j} d\omega_{j}}
{(2\pi)} F(\vec{K}_{1}-\vec{K}_{4}) \delta(\omega_{1}+\omega_{2} -
\omega_{3}-\omega_{4}) \\ 
\delta(\vec{K}_{1}+\vec{K}_{2}-\vec{K}_{3} - 
\vec{K}_{4}) \Psi_{\alpha}^{\dagger} (\vec{K}_{4}, \omega_{4}) 
\Psi_{\alpha}^{\dagger} (\vec{K}_{3}, \omega_{3}) 
\Psi_{\alpha}(\vec{K}_{2},\omega_{2})
 \Psi_{\alpha}(\vec{K}_{1},\omega_{1}), 
\end{eqnarray*}
and
\begin{eqnarray*}
 S_{dis}= \sum_{\alpha,\beta=1}^{N} 
\int (\prod_{m=1}^{4} 
d^{d}\vec{K}_{m}) 
\frac{d\omega_{1} d\omega_{2}}{(2\pi)^{2}} W(\vec{K}_{1}-\vec{K}_{3}) 
\delta(\vec{K}_{1}+\vec{K}_{2}-\vec{K}_{3}-\vec{K}_{4}) \\
\Psi_{\alpha}^{\dagger}(\vec{K}_{4}, \omega_{2}) 
\Psi_{\beta}^{\dagger}(\vec{K}_{3},
\omega_{1}) \Psi_{\alpha}(\vec{K}_{2}, \omega_{2}) 
\Psi_{\beta} (\vec{K}_{1}, \omega
_{1}).  
\end{eqnarray*}
Frequency integrals run over the whole real axis and the 
measure in $\vec{K}$-space in $d$-dimensions is defined as:
\begin{equation} 
\int d^{d}\vec{K}=\frac{\pi^{(d-3)/2}}{\Gamma(\frac{d-1}{2})} \int_{-\Lambda}
^{\Lambda} dk \int_{0}^{\pi}(\sin\theta)^{d-2} d\theta, 
\end{equation}
where $e(\vec{K})=k=|\vec{K}|-K_{F}$, $\Gamma(x)$ is the factorial 
function, and we assume a circular Fermi surface
 and the ultraviolet cutoff 
$\Lambda<< K_{F}$. We set $K_{F}=m=\hbar=1$.
Greek indices enumerate replicas and the 
zero-temperature free energy is $G=\int_{\Psi} (\exp(-S) -1)/N$, in 
the limit of vanishing number of replicas $N\rightarrow 0$. Note that the 
individual frequencies, and not only their sum, 
are conserved in $S_{dis}$, 
since the scattering over random impurities is elastic. 
This results in coupling function $W$ having an engineering scaling like 
frequency (or momentum) 
while $F$ is dimensionless \cite{shankar}, which indicates 
a finite scattering rate of the plane waves. 
We assume that the 
incoming (and consequently, the outgoing) 
momenta in both $S_{dis}$ and $S_{int}$ are always opposite in 
direction, since precisely 
these interactions are responsible for superconductivity \cite{shankar}
and weak-localization \cite{abrahams}, 
\cite{herbut}, when present separately. 
Both $F$ and $W$ essentially depend only on angle $\theta$ between the 
momenta in their arguments, due to the restriction on all momenta to 
the vicinity of the Fermi surface, and $F(\theta)=-F(\theta+\pi)$. 

 When $d\rightarrow 1$, the theory becomes one-dimensional  
since  $\int d^{d}\vec{K}\rightarrow \int_{-\Lambda}
^{\Lambda}  (dk/2\pi) \int d\theta (\delta(\theta) + \delta(\theta-\pi))$.
 At $d=1$, 
the interaction is described by a single coupling constant 
$g\equiv 2(F(\theta=0)-F(\theta=\pi))/\pi$, 
while disorder is determined by two coupling constants: 
$w_{b}\equiv 2 W(\theta=\pi)$ 
for the  backward, and $w_{f}\equiv 2 W(\theta=0)$ 
for the forward 
scattering over the random potential. Hereafter we set $w_{f}=0$, since 
forward scattering is unrelated to localization in $d=1$
\cite{remark} and drop 
the subscript "b" in $w_{b}$.  
In $d=1+\epsilon$, $0<\epsilon<<1$, for $F(\theta)$ and $W(\theta)$ smooth 
around $\theta=\pi$ the theory is still parametrized 
with only 
two coupling constants $g$ and $w$, as defined above. 
To the lowest order in 
$\epsilon$ and disorder, $\beta$-functions then assume a general form:
\begin{equation}
\beta_{g}\equiv \frac{d g}{d\xi}=-\epsilon A(g) + 
\hat{w} B(g)+O(\epsilon^2,\hat{w}^2, 
\epsilon \hat{w}), 
\end{equation}
\begin{equation}
\beta_{w}\equiv \frac{d \hat{w}}{d\xi}= 
\hat{w} C(g) +O(\epsilon^2,\hat{w}^2, \epsilon \hat{w}), 
\end{equation}
where $\hat{w}=(w/k)>0$ is the dimensionless disorder-variable, 
and $\xi=\ln(\Lambda/k)=\ln(\Lambda/\omega)$, where $k$ ($\omega$) 
measures the momentum (frequency) of the external legs in the 
renormalized disorder (interaction) vertex. 
We calculate the requisite functions using the standard diagrammatic 
perturbation theory in $d=1$,  to the second order 
in $g$: 
\begin{equation}
A(g)=\frac{g^2 \ln 2}{2} +O(g^3),
\end{equation}
\begin{equation}
B(g)=-\frac{1}{2} g + 0.10 g^2 + O(g^3),
\end{equation}
\begin{equation}
C(g)=1+g-\frac{1}{2} g^2 +O(g^3). 
\end{equation} 
Function $A(g)$ is positive, so at 
$\hat{w}=0$, in the infrared limit ($\xi\rightarrow\infty$)
$g\rightarrow -\infty$ if initially negative, indicating 
the opening of a gap in spectrum and 
the long-range superconducting order at $T=0$. 
The term 
proportional to $\epsilon$ in 
$\beta_{g}$ comes from the increase of the phase space available 
for the particle-particle, relative to the particle-hole, scattering 
when $\epsilon>0$, {\it and not}, as usually, from $g$ acquiring a 
finite engineering dimension. Recall that in  two and three dimensions 
diagrams of particle-hole type are suppressed by powers of 
$\Lambda/K_F$
due to the 
restriction that all scattering processes happen 
in the vicinity of the Fermi surface \cite{shankar}. 
In $d=1$ 
the Fermi surface is reduced to two points, and 
one-loop particle-particle and 
particle-hole diagrams are precisely 
equal and opposite in sign, so they cancel out
exactly. This cancelation persists to all orders in perturbation 
theory \cite{castro}, so $\beta_g \equiv 0$ in $d=1$ without disorder. 
For infinitesimal $\epsilon$ however, the homogenous 
system at $T=0$ may be expected to 
behaves qualitatively like in higher dimensions, and our lowest-order 
result for $A(g)$ conforms to this expectation. The second term 
in $\beta_{g}$ is 
positive for $g<0$, up to two-loops for $B(g)$. This is 
a pair-breaking property of the disorder, which therefore  
always inhibits superconductivity in this model. 
Also, $C(g)$ becomes reduced 
for $g<0$, indicating a suppression of localization by the 
attractive interaction in turn. The disorder-variable  
$\hat{w}$ scales precisely like 
the dimensionless resistance in $d=1$, and the 
eq. 5 for $g=0$ reduces to the statement of the Ohms  law. 
In $d=1+\epsilon$ the resistance and $\hat{w}$ should be expected 
to be related as 
$R \propto \hat{w} k^\epsilon$, but this distinction between 
the two variables  is of order 
of $\epsilon^2$ in the $\beta$-functions. Staying infinitesimally 
close to $d=1$ we thus may still identify 
$\hat{w}$ as the resistance, i. e. the right scaling variable. 
It is also interesting to note that 
the second-order (two-loop)
corrections in $B(g)$ and $C(g)$ produce only small quantitative, 
but no qualitative changes. 

When $\epsilon=0$ the problem has been studied before in refs. 7 and 8 
within the bosonized version of the theory (2) via 
mapping onto the equivalent system of 
two-dimensional classical Coulomb plasma. Our perturbative results 
are in general agreement with these calculations. 
Some numerical differences originate in 
different implementation of the renormalization procedure: 
the coefficient of the $g^2$-term in $C(g)$ is different, for 
example. 
The topology of the  scaling diagram, 
 the behavior of the conductivity 
at the transition and the divergence of the correlation length 
are the same as in 
our analysis at $\epsilon=0$, these 
differences notwithstanding. 

The main new feature brought by  finite 
 $\epsilon$ is the appearance of the 
finite-disorder fixed point in the physical region of the parameter space 
that determines the critical behavior at the 
quantum SI transition (Figure 1). It is located at 
\begin{equation}
\hat{w}_{c}=\frac{\epsilon A(g_{c})}{B(g_c)}\approx 0.42 \epsilon, 
\end{equation}
\begin{equation}
g_c \approx 1-\sqrt{3}, 
\end{equation}
where the numbers are obtained on the basis of the perturbative 
series in eqs. 6-8. 
Note that the fixed point is 
infinitesimal only in disorder and not in interaction, 
so our results are not exact even to the lowest order in $\epsilon$. 
Interaction at the fixed-point is of order $d_{\tau}=1$, the number of 
dimensions in directions of imaginary time. 
A more precise determination of the scaling equations 
would require a calculation and resummation 
of higher order terms in $g$. Under the assumption that there are no 
intermediate fixed points, scaling trajectories 
starting above the separatrix on 
Figure 1 end in the 
non-interacting, strongly disordered sink, which we 
identify with Anderson insulator. This should be contrasted with the 
Bose-glass phase in which tightly bound pairs of fermions would be 
localized. 
For initially positive interaction $g$, the  scaling is always towards 
the insulating phase, unless $\hat{w}=0$, when flows terminate 
in the trivial, 
unstable fixed point at the origin. 

The relevant direction 
at the SI fixed point plays the 
role analogous to temperature in thermal phase transitions. The 
linearization of the flow in the vicinity of the fixed point yields 
the correlation length exponent, $L_{cor}\propto \delta^{-\nu}$:
\begin{equation}
\nu=\frac{1}{\sqrt{\epsilon A(g_c) C'(g_{c})}}\approx 
\frac{1.75}{\sqrt{\epsilon}}, 
\end{equation}
where the prime denotes the derivative of the function and 
the parameter $\delta$ measures 
the distance from the transition. 
When $\epsilon\rightarrow 0$ 
the exponent  $\nu$ diverges, 
indicating the vicinity of the lower critical 
dimension for the transition. 
The value of the 
exponent satisfies the Harris criterion $\nu>2/d$ 
\cite{harris} for all values of $\epsilon$. 
To the lowest order in disorder and interaction, the dynamical 
critical exponent acquires a positive correction: 
\begin{equation}
z=1+\frac{\hat{w}_{c}}{2}+O(\hat{w}_{c} g_c, \hat{w}_c^2),  
\end{equation}
as a result of different treatment of frequences and momenta 
by the disorder term in the action (2), 
which breaks the relativistic invariance 
of the homogeneous system.

We may define  the superconducting quasi-transition temperature $T_{sc}$ 
(proportional to the superconducting gap at $T=0$), as 
the frequency (momentum) scale 
at which $g$ diverges. When $\hat{w}=0$, it 
follows from the eq. 4 that: 
\begin{equation}
T_{0,sc}=\Lambda \exp{-\frac{2}{\epsilon |g_{0}| \ln 2}}, 
\end{equation}
where $g_0<0$ is the initial value of $g$. 
True superconducting temperature is of 
course zero for $d<2$ due to the effect of thermal fluctuations which 
are not included properly in our zero-temperature theory, nevertheless, 
the above defined quantity sets the temperature 
scale at which the Cooper pairs start to form. 
On the superconducting side, close to the SI transition this 
energy scales to zero as 
$T_{sc}\propto \delta ^{z\nu}$. 
The diverging size of the Cooper pairs at the transition is a 
consequence of the pair-breaking nature of the disorder in our model.
It thus seems unlikely that the SI transition in this systems 
can be described by an effective bosonic theory in 
terms of preformed Cooper pairs.

Under the assumption that at low, but finite temperatures,  
 scaling equations do 
not get substantially modified,
 we may infer the temperature dependence of the 
conductivity \cite{giamarchi}. At the superconducting side of the 
transition, $\hat{w}\rightarrow 0$, and 
the conductivity is given by  the sum of classical (Boltzmann) part and 
quantum (weak-localization) correction: $\sigma=\sigma_{B}
+\sigma_{Q}$. According to our calculations  
$\hat{w}$ vanishes at some finite temperature 
$T_{sc}$, at which $g$ diverges. At this temperature superconducting 
fluctuations become strong which results in dramatic reduction in 
resistivity.
Within our approximation, $\sigma_{B}$ actually diverges at $T_{sc}$,  
while $\sigma_{Q}\sim - L_T ^{1-\epsilon}$, where 
$L_T \propto T^{-1}$ is the inelastic scattering length,  remains finite. 
Thus $\sigma=\sigma_{B}$ close to $T_{sc}$, 
and in our units $\sigma_{B}=1/(\hat{w} k^{1-\epsilon})$, 
with  temperature 
scaling as momentum 
$T=k$. The behavior of $\hat{w}$ close to $T_{sc}$ is determined by the 
scaling 
equations in the region $g<<-1$. From the eq. (8) $C(g)\approx -g^{2}/2$
in this region, implying the resistivity, to the lowest order in 
$\epsilon$:
\begin{equation}
\rho\equiv\sigma^{-1}\approx \hat{w}_{0} T
\exp{ -\frac{2 T_{sc}}{ \epsilon^{2} (\ln2)^2 (T-T_{sc})}}. 
\end{equation}
For more 
general $A(g)$ and $C(g)$, we would still expect to find an activated form 
of the resistivity close to $T_{sc}$, 
but with the power of $(T-T_{sc})$ and the constants in the exponent 
modified.

On the insulating side of the transition, 
at low temperatures the system is in the 
localized phase and the transport proceeds via hopping between the localized 
states. The conductivity then should have an activated form, similarly (but 
not identically) to the resistivity in the superconducting phase close to 
$T_{sc}$:  
\begin{equation}
\sigma\sim \exp{ -(\frac{T_{loc}}{T})^\alpha}, 
\end{equation}
where the power $\alpha$ depends on the hopping mechanism. 
The temperature scale $T_{loc}\sim 1/L_{loc}^z$, where $L_{loc}$ is the 
localization length at the insulating side. This length may be defined 
as a scale at which $\hat{w}$ becomes of order of unity upon scaling, so  
$L_{loc}\propto L_{cor}$ and $T_{loc}$ also vanishes as
$T_{loc}\propto \delta^{z\nu}$. 
In $d=1$, $L_{loc}\propto\exp(const/\delta^{1/2})$, 
akin to the 
divergence of correlation length 
in the Kosterlitz-Thouless transition \cite{giamarchi}. 

Right at the SI transition the resistance $\hat{w}$ has a finite 
value $\hat{w}_{c}\propto\epsilon$. The 
conductivity is still determined by 
the weak-disorder expression. Now 
$\sigma_{B}=T^{-(1-\epsilon)/z} /\hat{w}_{c}$ as $T\rightarrow 0$, and the 
quantum correction $\sigma_{Q}\sim T^{-(1-\epsilon)/z}$ 
is still negligible infinitesimally close to $d=1$. 
The conductivity at the 
point of SI transition thus diverges as:  
\begin{equation}
\sigma=\frac{T^{-(1-const. \epsilon)}}{\hat{w}_{c}},  
\end{equation}
with $const.\approx 1.20$, to the lowest order in $g$.
When $\epsilon=0$, we find $\sigma=\sigma_{B}\propto(\ln T)^2 /T$ 
right at the transition. 

The results presented in this paper 
are not directly relevant for the experiments on thin 
films \cite{haviland}
due to the assumed spinless nature of fermions in our model. 
Disorder is not pair-breaking for the usual s-wave pairing, and the 
SI transition in that case should be in a different universality 
class, for which the theory in terms of the 
bosonic Cooper pairs becomes viable 
\cite{wallin}. 
The transition between a p-wave 
superconductor and insulator, however, could be expected 
to be in the same universality class as the considered model. 
$^3 He$ in strongly disordered media \cite{trey} 
such are aerogels may offer an experimental realization of the quantum  
transition we considered. 

In summary, we studied a $1+\epsilon$-dimensional model of 
interacting, disordered fermions which exhibits a direct, continuous, 
superconductor-Anderson insulator transition at zero temperature. 
The value of 
parameter $\epsilon$ controls the value of resistance at the quantum 
fixed point, and we calculated the critical exponents to the lowest 
order in $\epsilon$ and interaction. The behavior of conductivity at 
both sides of the transition is obtained, and the superconducting gap 
at the critical point is showed to vanish. The model may have 
relevance  for p-wave superconductors in disordered media at low 
temperatures. 

We acknowledge useful conversations with Professors I. Affleck, 
P. Stamp and Z. Te\v sano-vi\' c. The author thanks NSERC of Canada 
and Izaak Walton Killam foundation for financial support. 
\vspace{10mm}


\vspace{10mm}

\noindent
Figure Captions:
\hspace{6mm}

\noindent
Figure 1: Schematic scaling diagram in the interaction-disorder plane 
in $d=1+\epsilon$ for spinless fermions. The scaling trajectory flowing 
right into the fixed point separates the flows towards the superconducting 
($g=-\infty, \hat{w}=0$) phase from the flows towards the 
non-interacting localized insulator 
($g=0, \hat{w}=\infty$).

\pagebreak


\begin{thebibliography}{99}
\bibitem{abrahams} E. Abrahams, P. W. Anderson, D. C. 
Liccardello and T. V. Ramakrishnan, Phys. Rev. Lett. 
{\bf 42}, 673 (1979)
\bibitem{haviland} For review, see 
Y. Liu and A. Goldman, Mod. Phys. Lett B, 
{\bf 8}, 277 (1994)
\bibitem{weichman} M. P. A. Fisher, P. B. Weichman, G. Grinstein and 
D. S. Fisher, Phys. Rev. B {\bf 40}, 546 (1989)
\bibitem{fisher} M. P. A. Fisher, Phys. Rev. Lett. {\bf 65}, 923 (1990)
\bibitem{wallin} M. Wallin, E. S. Sorensen, S. M. Girvin and 
A. P. Young, Phys. Rev. B, {\bf 49}, 12115 (1994) and references 
therein.
\bibitem{halperin} M. Ma, B. I. Halperin and P. A. Lee, Phys. Rev. B 
{\bf 34},
3136 (1986)
\bibitem{apel} W. Apel, J. Phys. C: Solid State Phys. {\bf 15}, 1973 (1982)
\bibitem{giamarchi} T. Giamarchi and H. J. Shulz, Phys. Rev. B {\bf 37}, 327 
(1988)
\bibitem{solyom} J. Solyom, Adv. in Phys. {\bf 28}, 201 (1979)
\bibitem{edwards} S. Edwards and P. W. Anderson, J. Phys. F {\bf 5}, 
965 (1975)
\bibitem{shankar} R. Shankar, Rev. Mod. Phys. {\bf 66}, 129 (1994)
\bibitem{herbut} I. F. Herbut, unpublished
\bibitem{remark} Formally, this is manifested in cancelation of 
terms containing $w_{f}$ in 
$\beta$-functions for $g$ and $w_{b}$ in $d=1$. See also, A. A. Abrikosov 
and J. A. Ryzhkin, Adv. Phys. {\bf 27}, 147 (1978)
\bibitem{castro} C. Di Castro and W. Metzner, Phys. Rev. Lett. {\bf 67}, 
3852 (1991)
\bibitem{harris} A. B. Harris, J. Phys. C: Solid State Phys. {\bf 7}, 
1671 (1974); J. Chayes et al., Phys. Rev. Lett. {\bf 57}, 2999 (1986)
\bibitem{trey} J. V. Porto and J. M. Parpia, Phys. Rev. Lett. {\bf 74}, 
4667 (1995)
\end{thebibliography}
\end{document}